\documentclass[aps,prb,twocolumn,showpacs,floatfix,superscriptaddress]{revtex4}
\usepackage{amsfonts}
\usepackage{amssymb}
\usepackage{amsmath}
\usepackage{graphicx}
\usepackage{dcolumn}
\usepackage{bm}

\begin{document}

\title{One-dimension cubic--quintic Gross--Pitaevskii equation in
Bose-Einstein condensates in a trap potential}
\author{C. Trallero-Giner}
\affiliation{Centro Brasileiro de Pesquisas Fisicas, Rua Xavier Sigaud 150, 22290-180 Rio
de Janeiro-RJ, Brazil}
\author{R. Cipolatti}
\affiliation{Instituto de Matem\'{a}tica, Universidade Federal do Rio de Janeiro, C.P.
68530, Rio de Janeiro, RJ, Brasil}
\date{\today }

\begin{abstract}
By means of new general variational method we report a direct solution for
the quintic self-focusing nonlinearity and cubic-quintic 1D Gross Pitaeskii
equation (GPE) in a harmonic confined potential. We explore the influence of
the 3D transversal motion generating a quintic nonlinear term on the ideal
1D pure cigar-like shape model for the attractive and repulsive atom-atom
interaction in Bose Einstein condensates (BEC). Also, we offer a closed
analytical expression for the evaluation of the error produced when solely
the cubic nonlinear GPE is considered for the description of 1D BEC.
\end{abstract}

\maketitle

\section{Introduction}

Nowadays one and quasi-one dimensional Bose-Einstein condensates (BEC) are
common experimental procedures~\cite{Gor}. The transition from 3D to 1D
system was invoked long time ago~\cite{lieb1}. In general, the 3D
Gross-Pitaesvkii equation (GPE) cannot be factorized into transverse and
longitudinal motions, nevertheless, under certain parameter regions we can
assert that the BEC follows a 1D behavior (for a detailed discussion see
Ref.~\onlinecite{lieb2}). In the case of the harmonic trapping potential and
considering that the atoms are tightly confined in two transverse
directions,\ a transition to the quasi-1D description is possible. Starting
with the standard 3D GPE, employing the adiabatic approximation and using
the anzat wavefunction $\Psi (x,\mathbf{r;}t)=\exp (-i\mu _{0}t/\hslash
)\Phi (x)\chi (\mathbf{r;}t)$, we can derive an effective 1D GPE, which
describes the physical characteristics of the cigar-like shape condensate~
\cite{Carretero,coefic,Khaykovich}

\begin{equation}
-\frac{\hbar ^{2}}{2m}\frac{d^{2}\Phi }{dx^{2}}+\frac{1}{2}m\omega
^{2}x^{2}\Phi +g_{_{1D}}\left\vert \Phi \right\vert ^{2}\Phi -g\left\vert
\Phi \right\vert ^{4}\Phi =\mu _{0}\Phi ,  \label{CQ}
\end{equation}
where $\mu _{0}\in {\mathbb{R}}$ is the chemical potential, $m$ is the
atomic mass, $\omega $ is the longitudinal harmonic oscillator frequencies,
and $g_{1D}$ $\in {\mathbb{R}}$, $g\in {\mathbb{R}}$ are the effective 1D
nonlinear self-interaction coefficients. These two coefficients depend on
the total number $N$ of particles in the condensate, the transverse harmonic
oscillator frequency $\omega _{\mathbf{r}}$ and the scattering length $a_{s}$
($a_{s}>0$ or $a_{s}<0$ for attractive or repulsive interatomic interaction,
respectively) by the relations $g_{_{1D}}=2a_{s}N\hbar \omega _{\mathbf{r}}$
and $g=6\ln (4/3)g_{_{1D}}^{2}/\hbar \omega _{\mathbf{r}}$~\cite{coefic},
where we have chosen for the stationary state $\Phi $ the normalization
condition $\int\limits_{{\mathbb{R}}}dx\left\vert \Phi \right\vert ^{2}=1.$

Equation~(\ref{CQ}) is a cubic-quintic nonlinear Sch\"{o}dinger equation
(NLSE) with real coefficients. The presence of the $-g\left\vert \Phi
\right\vert ^{4}$ term in (\ref{CQ}) is due to the deviation from one
dimension on the longitudinal condensate dynamics, i.e.\ a residual three
dimensionality\emph{\ }on an effective one-dimensional GPE. In the case of a
homogeneous\ medium, i.e. assuming that $\omega $ is zero, the cubic-quintic
NLSE was widely used to describe\ the physical process of an optical medium
with a nonlinear polarization including susceptibilities up to fifth order
\cite{Pushkarov}. Also, the soliton solutions have been extensively studies
in Refs.~\onlinecite{coefic,Khaykovich,Sinha}.

As we have mentioned above, Eq.~(\ref{CQ}) is the cigar-like shape approach
from the 3D GPE. An important issue is the range of validity of (\ref{CQ}),
which is directly linked to the existence and stability of set of ground
states solutions of the 3D NLSE. It is well known that for any value of $
a_{s}>0$ the 3D GPE does not collapse~\cite{Carretero}. However, for
attractive interatomic interactions, the solution is dynamically stable if
and only if $a_{s}$ is within the range~\cite{Khaykovich,Perez}
\begin{equation}
\frac{N|a_{s}|}{a_{\perp }}<0.627,  \label{const}
\end{equation}
with $a_{\perp }=\sqrt{\hbar /m\omega _{\mathbf{r}}}$. Hence, the validity
of the cigar-like shape approach represented by Eq.~(\ref{CQ}) is also
restricted to the constrain (\ref{const}).

Rescaling to dimensionless variables $l_{o}=\sqrt{\hbar /m\omega }$, $\xi
=x/l_{o},$ $\lambda =2g_{_{1D}}/(l_{o}\hbar \omega ),$ $\mu =2\mu
_{0}/(\hbar \omega ),$ $\psi (\xi )=\Phi (\xi l_{o})/\sqrt{l_{o}},$ $
\varepsilon =3\ln (4/3)\omega /\omega _{\mathbf{r}},$ Eq.~(\ref{CQ}) can be
cast as

\begin{equation}
-\frac{d^{2}\psi }{d\xi ^{2}}+\xi ^{2}\psi +\lambda \left\vert \psi
\right\vert ^{2}\psi -\varepsilon \lambda ^{2}\left\vert \psi \right\vert
^{4}\psi =\mu \psi .  \label{CQAd}
\end{equation}

The main task of this paper is the implementation of a more general
variational mathematical approach to solve Eq.~(\ref{CQAd}). Based on this
result, we provide approximate solutions for the order parameter, the
chemical potential and minimal energy for the quintic and cubic-quintic
GPEs. The paper is organized as follows. First, in Sec. II we present the
bases of our formalism, i.e., we present exact formulae for the energy and
the chemical potential as functions of relevant parameters of (\ref{CQAd}).
By considering a trial function for the ground state, we derive in Sec. III
a representation for the energy, $E_{app}(\lambda ),$ and chemical
potential, $\mu _{app}(\lambda )$. Section IV is devoted to the application
of our results to get explicit approximate solutions for the quintic and
also cubic-quintic NLSEs. An estimation of the error due to the influence of
the interaction between the axial and radial degrees of freedom on the 1D
cigar-shape model is presented both graphically and analytically as function
of the self-interaction parameter $\lambda $ and the coefficient $
\varepsilon $ leading the quintic nonlinear term.

\section{Cubic-quintic nonlinear Gross-Pitaevskii equation}

In the following we will consider a more general nonlinear Gross-Pitaevskii
equation

\begin{equation}
-\frac{d^{2}\psi }{d\xi ^{2}}+\xi ^{2}\psi +a\lambda \left\vert \psi
\right\vert ^{2}\psi -b\lambda ^{2}\left\vert \psi \right\vert ^{4}\psi =\mu
\psi ,\text{ \ \ }\xi \in {\mathbb{R}}\mathbf{,}  \label{CQAdG}
\end{equation}
where $a\geq 0$ and $b$ are real constants.

Let $\mathbf{V}=\left\{ \psi \in H^{1}({\mathbb{R}})|\text{ }\int\limits_{{\
\mathbb{R}}}\xi ^{2}\left\vert \psi (\xi )\right\vert ^{2}d\xi <+\infty
\right\} $ be the Hilbert space endowed~\cite{rolci} with the norm

\begin{equation*}
\left\Vert \psi \right\Vert _{\mathbf{V}}=\left[ \int\limits_{{\mathbb{R}}
}\left\vert \psi (\xi )\right\vert ^{2}d\xi +\int\limits_{{\mathbb{R}}}\xi
^{2}\left\vert \psi (\xi )\right\vert ^{2}d\xi \right] ^{1/2}
\end{equation*}
and the corresponding inner product

\begin{equation*}
\left( \phi |\psi \right) _{\mathbf{V}}=\int\limits_{{\mathbb{R}}}\left(
\frac{d\phi }{d\xi }\frac{d\psi }{d\xi }+\xi ^{2}\phi (\xi )\psi (\xi
)\right) d\xi .
\end{equation*}
In $\mathbf{V}$ we define the energy functional
\begin{eqnarray}
E_{\lambda }[\psi ] &=&\int\limits_{{\mathbb{R}}}\left\vert \frac{d}{d\xi }
\psi (\xi )\right\vert ^{2}d\xi +\int\limits_{{\mathbb{R}}}\xi
^{2}\left\vert \psi (\xi )\right\vert ^{2}d\xi  \notag \\
&&+\frac{a\lambda }{2}\int\limits_{{\mathbb{R}}}\left\vert \psi (\xi
)\right\vert ^{4}d\xi -\frac{b\lambda ^{2}}{3}\int\limits_{{\mathbb{R}}
}\left\vert \psi (\xi )\right\vert ^{6}d\xi .  \label{Funtional}
\end{eqnarray}
We denote by $\mathcal{G}_{\lambda }$ the set of ground states of Eq.~(\ref
{CQAdG}), i.e., the set of functions of $\mathbf{V}$ that minimize the
energy functional $E_{\lambda }[\psi ]$ under the condition

\begin{equation}
Q[\psi ]=\int\limits_{{\mathbb{R}}}\left\vert \psi (\xi )\right\vert
^{2}d\xi =1.  \label{CondNo}
\end{equation}
Notice that in the case of attractive interaction where $b>0$, it is
possible to show (by applying the Gagliardo-Nirenberg inequalities) that the
set of ground states $\mathcal{G}_{\lambda }$ is nonempty if the condition $
\left\vert \lambda \right\vert <2/\sqrt{|b|}$ is satisfied. Hence, for $\psi
_{\lambda }\in $ $\mathcal{G}_{\lambda }$ we obtain

\begin{equation}
\frac{dE_{\lambda }[\psi _{\lambda }]}{d\lambda }=\left\langle \frac{\delta
E_{\lambda }[\psi _{\lambda }]}{\delta \psi _{\lambda }};\frac{d\psi
_{\lambda }}{d\lambda }\right\rangle +\frac{a}{2}\left\Vert \psi _{\lambda
}\right\Vert _{4}^{4}-\frac{2b\lambda }{3}\left\Vert \psi _{\lambda
}\right\Vert _{6}^{6},  \label{DerEmin}
\end{equation}
where

\begin{equation}
\left\Vert \psi (\xi )\right\Vert _{4}^{4}=\int\limits_{{\mathbb{R}}
}\left\vert \psi (\xi )\right\vert ^{4}d\xi \text{ \ and \ }\left\Vert \psi
(\xi )\right\Vert _{6}^{6}=\int\limits_{{\mathbb{R}}}\left\vert \psi (\xi
)\right\vert ^{6}d\xi  \label{norma4}
\end{equation}
are the usual norms of the standard Banach spaces $L^{4}({\mathbb{R}})$ and $
L^{6}({\mathbb{R}})$. Taking into account the relation between the energy
and the chemical potential, i.e. $\delta E_{\lambda }[\psi _{\lambda
}]/\delta \psi _{\lambda }=\mu \psi _{\lambda }$, it follows that

\begin{eqnarray}
\left\langle \frac{\delta E_{\lambda }[\psi _{\lambda }]}{\delta \psi
_{\lambda }};\frac{d\psi _{\lambda }}{d\lambda }\right\rangle &=&\mu
\left\langle \psi _{\lambda };\frac{d\psi _{\lambda }}{d\lambda }
\right\rangle  \notag \\
&=&\frac{\mu }{2}\frac{\delta Q_{\lambda }[\psi _{\lambda }]}{\delta \psi
_{\lambda }}=0.  \label{Cond}
\end{eqnarray}
Hence, from Eqs.~(\ref{DerEmin}) and (\ref{Cond}) we obtain the useful
formula

\begin{equation*}
\frac{dE_{\lambda }[\psi _{\lambda }]}{d\lambda }=\frac{a}{2}\left\Vert \psi
_{\lambda }\right\Vert _{4}^{4}-\frac{2b\lambda }{3}\left\Vert \psi
_{\lambda }\right\Vert _{6}^{6}.
\end{equation*}
Thus, the minimum energy is given by

\begin{equation}
E_{\min }(\lambda )=1+\frac{a}{2}\int\limits_{\mathbf{0}}^{\lambda
}\left\Vert \psi _{s}\right\Vert _{4}^{4}ds-\frac{2b}{3}\int\limits_{\mathbf{
\ \ \ 0}}^{\lambda }s\left\Vert \psi _{s}\right\Vert _{6}^{6}ds.
\label{Emin}
\end{equation}
Moreover, from Eq.~(\ref{CQAd}) is straightforward that the chemical
potential can be written as

\begin{equation*}
\mu _{\min }(\lambda )=E_{\min }(\lambda )+\frac{a\lambda }{2}\left\Vert
\psi _{\lambda }\right\Vert _{4}^{4}-\frac{2b\lambda ^{2}}{3}\left\Vert \psi
_{\lambda }\right\Vert _{6}^{6}.
\end{equation*}
or equivalently~\cite{rolci}

\begin{eqnarray}
\mu _{\min }(\lambda ) &=&1+\frac{a}{2}\left[ \lambda \left\Vert \psi
_{\lambda }\right\Vert _{4}^{4}+\int\limits_{\mathbf{0}}^{\lambda
}\left\Vert \psi _{s}\right\Vert _{4}^{4}ds\right]  \notag \\
&&-\frac{2b}{3}\left[ \lambda ^{2}\left\Vert \psi _{\lambda }\right\Vert
_{6}^{6}+\int\limits_{\mathbf{0}}^{\lambda }s\left\Vert \psi _{s}\right\Vert
_{6}^{6}ds\right] .  \label{mumin}
\end{eqnarray}
It is important to remark that Eqs.~(\ref{Emin}) and (\ref{mumin}) are exact
under the condition of knowing the ground state $\psi _{\lambda }\in $ $
\mathcal{G}_{\lambda }$ and therefore, independent of the method or approach
we employe to get the solution of the Eq.~(\ref{CQAd}).

\section{Approximate formulae}

It is possible to show~\cite{rolci} that any solution of Eq.~(\ref{CQAdG})
belonging to the Hilbert space $\mathbf{V}$ has the asymptotic behavior $
\exp (-\tau \xi ^{2}),$ with $\tau >0,$ as $|\xi |\rightarrow \infty $. So,
to evaluate the minimal energy $E_{\min }(_{\lambda }),$ we can consider for
the ground state the trial function

\begin{equation}
\psi _{\tau }(\xi )=\left( \frac{2\tau }{\pi }\right) ^{1/4}\exp (-\tau \xi
^{2}).  \label{trialf}
\end{equation}
Using the function (\ref{trialf}) and evaluating the energy functional (\ref
{Funtional}), we obtain the algebraic expression

\begin{equation}
E_{\lambda }[\psi _{\tau }]=(1-\epsilon \lambda ^{2})\sigma ^{2}+\frac{1}{
4\sigma ^{2}}+\frac{a\lambda }{2\sqrt{\pi }}\sigma   \label{Efitau}
\end{equation}%
with $\epsilon =2b/(3\sqrt{3}\pi )$ and $\sigma =\sqrt{\tau }.$ In the case $
\epsilon \leq 0$ the Eq.~(\ref{Efitau}) presents a global minimum in $
(0,+\infty )$ for any $a$ and $\lambda \in {\mathbb{R}}$, but for $\epsilon
>0$ a global minimum is guaranteed if $|\lambda |<1/\sqrt{\epsilon }=\sqrt{
3\pi \sqrt{3}}/\sqrt{2b}$.

Let $\sigma (\lambda )=\sqrt{\tau (\lambda )}$ be the minimizer of Eq.~(\ref
{Efitau}), in this way the ground state solution in $\mathcal{G}_{\lambda }$
can be searched considering the function $\varphi _{\lambda }(\xi ):=\psi
_{\tau (\lambda )}(\xi ).$ By using Eqs.~(\ref{Emin}) and (\ref{mumin}) we
obtain the approximate energy, $E_{app}(\lambda ),$ and chemical potential, $
\mu _{app}(\lambda ),$ namely
\begin{equation}
E_{app}(\lambda )=1+\frac{a}{2\sqrt{\pi }}\int\limits_{\mathbf{0}}^{\lambda
}\sigma (s)ds-2\epsilon \int\limits_{\mathbf{0}}^{\lambda }s\sigma ^{2}(s)ds
\label{Eapp}
\end{equation}
and

\begin{eqnarray}
\mu _{app}(\lambda ) &=&1+\frac{a}{2\sqrt{\pi }}\left[ \lambda \sigma
(\lambda )+\int\limits_{\mathbf{0}}^{\lambda }\sigma (s)ds\right]  \notag \\
&&-2\epsilon \left[ \lambda ^{2}\sigma ^{2}(\lambda )+\int\limits_{\mathbf{0}
}^{\lambda }s\sigma ^{2}(s)ds\right] .  \label{muapp}
\end{eqnarray}
It becomes clear that $|\lambda |$ $<2/\sqrt{b}<\lambda _{S}=\sqrt{3\pi
\sqrt{3}}/\sqrt{2b},$ ensuring that the set $\mathcal{G}_{\lambda }$ is
nonempty and that the functions $E_{app}(\lambda )$ and $\mu _{app}(\lambda
) $ can be good approximations of $E_{\min }(\lambda )$ and $\mu _{\min
}(\lambda )$, respectively.

\section{Applications and discussion of the results}

Explicit formulae (\ref{Emin}), (\ref{mumin}) and the approximate
relationships (\ref{Eapp}), (\ref{muapp}) are among the main results of our
work. Nevertheless, more explicit expressions for the order parameter $
\varphi _{\lambda },$ the energy and the chemical potential as a function of
the atom-atom interaction term would be desirable. To do so, let $\sigma
_{\min }=$ $\sigma _{\min }(a,\epsilon ;\lambda )$ be the global minimizer
of Eq.~(\ref{Efitau})$.$ Thus, Eqs.~(\ref{Eapp}) and (\ref{muapp}) with $
\sigma =\sigma _{\min }$ allow to obtain the dependence $E_{app}$ and $\mu
_{app}$ on the relevant physical parameters $a,\epsilon $ and $\lambda .$
Depending on the values of $\lambda $ and the sign of the parameter $
\epsilon $, we can distinguish several phases linked to both type of
interaction strengths, i.e. i) pure attractive ($\lambda <0,\epsilon >0)$,
ii) pure repulsive ($\lambda >0,\epsilon <0),$ and iii) a mixture ($\lambda
<0,\epsilon <0$ or $\lambda >0,\epsilon <0$).

More precisely, the value of $\tau _{\min }$ ($\sigma _{\min }(\lambda ):=
\sqrt{\tau _{\min }}$) that minimizes the function $\tau \longmapsto $ $
E_{\lambda }[\psi _{\tau }]$ fulfil the equation

\begin{equation}
(1-\epsilon \lambda ^{2})\sigma ^{4}+\frac{a\lambda }{4\sqrt{\pi }}\sigma
^{3}=\frac{1}{4}.  \label{trans}
\end{equation}
Notice that, for $\epsilon \leq 0$ or $\epsilon >0$ under the condition $
|\lambda |$ $<1/\sqrt{\epsilon },$ the function (\ref{Efitau}) is strictly
convex and coercive on the interval $(0,+\infty ),$ and in consequence Eq.~(\ref{trans}) has a unique solution, while for $|\lambda |>1/\sqrt{\epsilon }$
the function (\ref{Efitau}) is not bounded from below.

\subsection{Quintic NLSE}

Firstly and for sake of comparison, we will consider the quintic NLSE in a
harmonic potential~\cite{nota}. Choosing $a=0,$ from (\ref{trans}) we have

\begin{equation}
\sigma _{\varphi ^{5}}^{2}=\frac{1}{2\left( 1-\epsilon \lambda ^{2}\right)
^{1/2}}.  \label{Fi5}
\end{equation}
\begin{figure}[th]
\includegraphics[scale=0.30]{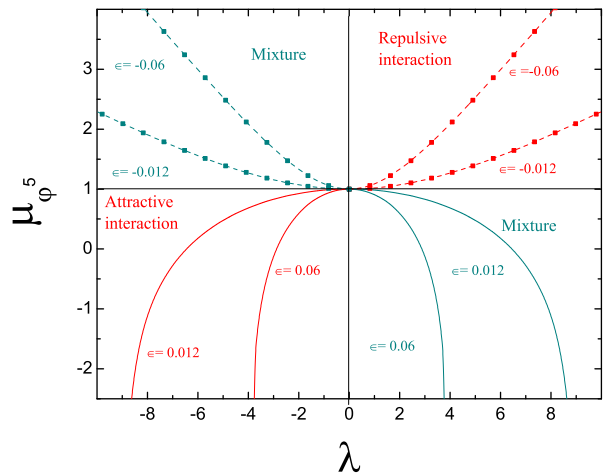}
\caption{(Color online) $\protect\mu _{\protect\varphi ^{5}}-\protect\lambda
$ map for the parameter values $\protect\epsilon =\pm 0.06$ and $\pm 0.012$.
According to the signs of $\protect\lambda $ and $\protect\epsilon $ the
character of the $\protect\mu _{\protect\varphi ^{5}}$, as approximated
solution of Eq.~(\protect\ref{CQAdG}) with $a=0$, can be mapped into fourth
zones: $\protect\lambda >0$, $\protect\epsilon <0$ -pure repulsive
interaction; $\protect\lambda <0$, $\protect\epsilon >0$ -pure attractive
interaction; $\protect\lambda >0$, $\protect\epsilon >0$ and $\protect
\lambda <0$, $\protect\epsilon <0$ -mixture region. }
\label{1}
\end{figure}
Inserting (\ref{Fi5}) into Eqs.~(\ref{Eapp}) and (\ref{muapp}) with $a=0$,
we obtain for the approximate energy $E_{\varphi ^{5}}$ and chemical
potential $\mu _{\varphi ^{5}}$ the expressions

\begin{equation}
E_{\varphi ^{5}}=\left( 1-\epsilon \lambda ^{2}\right) ^{1/2}  \label{Efi5}
\end{equation}
and

\begin{equation}
\mu _{\varphi ^{5}}=\left( 1-\epsilon \lambda ^{2}\right) ^{1/2}-\frac{
\epsilon \lambda ^{2}}{\left( 1-\epsilon \lambda ^{2}\right) ^{1/2}}.
\label{mufi5}
\end{equation}

Figure~\ref{1} displays the $\mu _{\varphi ^{5}}-$ $\lambda $ map diagram
for several values of the parameter $\epsilon .$ It can be seen that the
reduced chemical potential $\mu _{\varphi ^{5}}$=$2\mu _{0}/(\hbar \omega )$
shows a strong dispersion as a function of $\lambda $, moreover and
following the symmetry properties of Eq.~(\ref{mufi5}), we observe that $\mu
_{\varphi ^{5}}$ increases (decreases) for the pure repulsive phase, $
\lambda >0,$ $\epsilon <0$ (pure attractive phase, $\lambda <0,$ $\epsilon
>0 $), while in the mixture region an opposite behavior is reached with
respect\ to the pair of values ($\lambda ,\epsilon )$.

Following Eqs.~(\ref{trialf}) and (\ref{Fi5}) we obtain for the wavefunction
$\varphi _{\varphi ^{5},\lambda }$ the expression
\begin{equation}
\varphi _{\varphi ^{5},\lambda }(\xi )=\frac{1}{\left[ \pi \left( 1-\epsilon
\lambda ^{2}\right) ^{1/2}\right] ^{1/4}}\exp \left[ -\frac{\xi ^{2}}{
2\left( 1-\epsilon \lambda ^{2}\right) ^{1/2}}\right]  \label{WFi5}
\end{equation}
valid for $1>\lambda ^{2}\epsilon $. The above obtained wave function
exhibits different behavior depending on the sign of $\epsilon $ and
independent of the type interaction (attractive with $\lambda <0$ or
repulsive for $\lambda >0$). The function becomes effectively less confined
for $\epsilon <0$, i.e. $\varphi _{\varphi ^{5},\lambda }$ is delocalized
and its maximum decreases, while for $\epsilon >0$ the function $\varphi
_{\varphi ^{5},\lambda }(\xi )$ gets more localized and the maximum
increases as the nonlinear potential $\lambda ^{2}\epsilon $ increases.

\subsection{Cubic-quintic NLSE}

Although one can solve Eq.~(\ref{trans}) numerically and to obtain from
Eqs.~(\ref{Eapp}) and (\ref{muapp}) the energy $E_{\varphi ^{3}-\varphi
^{5}}(\lambda ,b)$ and the chemical potential $\mu _{\varphi ^{2}-\varphi
^{5}}(\lambda ,b),$ it will be very useful to report explicit compact
approximate solution of the cubic-quintic nonlinear 1D GPE (\ref{CQAdG}).
Searching the solution of Eq.~(\ref{trans}) with $a=1$ and $b=\varepsilon >0$
as a Taylor series on $\lambda ,$ $\sigma _{\varphi ^{3}-\varphi
^{5}}=\sum\limits_{n=0}^{\infty }\sigma _{n}\lambda ^{n}$, we get

\begin{eqnarray}
\sigma _{\varphi ^{3}-\varphi ^{5}} &=&\frac{1}{\sqrt{2}}-\frac{\lambda }{16
\sqrt{\pi }}+  \notag \\
&&\frac{\sqrt{2}}{2\pi }\left( \frac{3}{256}+\frac{\varepsilon }{6\sqrt{3}}
\right) \lambda ^{2}-  \notag \\
&&\frac{1}{2\pi \sqrt{\pi }}\left( \frac{1}{512}+\frac{\varepsilon }{12\sqrt{
3}}\right) \lambda ^{3}+  \notag \\
&&\frac{\sqrt{2}}{3\pi ^{2}2^{11}}\left( \frac{45}{128}+\frac{42\varepsilon
}{\sqrt{3}}\right) \lambda ^{4}+...  \label{sigma2-5}
\end{eqnarray}

\begin{figure}[h]
\includegraphics[scale=0.38]{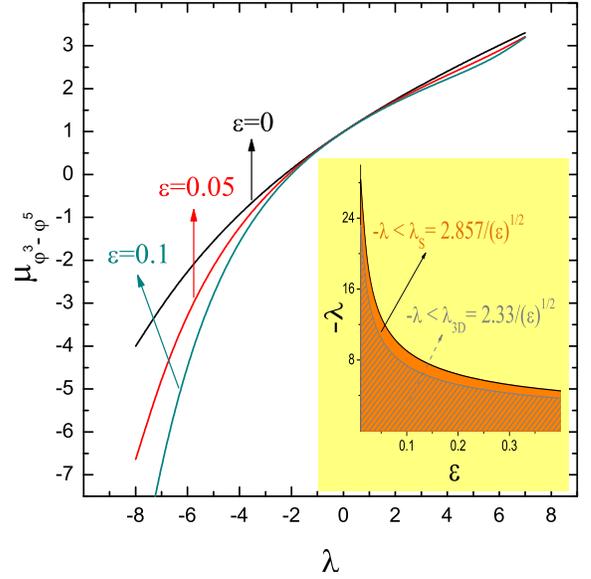}
\caption{(Color online) Reduced chemical potential $\protect\mu _{\protect
\varphi ^{3}-\protect\varphi ^{5}}$ as a function of the dimensionless
self-interaction parameter $\protect\lambda $, calculated from Eqs.~(\protect
\ref{Eng})-(\protect\ref{mufi2-5}) and several values of coefficient $
\protect\varepsilon $. Inset: $\protect\lambda $ - $\protect\varepsilon $
map showing the validity ranges of Eqs.~(\protect\ref{sigma2-5})-(\protect
\ref{mufi2-5}) and the stability region of 3D GPE given by $-\protect\lambda
<\protect\lambda _{S}=2.857/(\protect\varepsilon )^{1/2}$ (orange) and $-
\protect\lambda <\protect\lambda _{3D}=2.33/(\protect\varepsilon )^{1/2}$
(gray lines), respectively.}
\label{2}
\end{figure}
Under the condition $|\varepsilon |<<1$ and by substituting Eq.~(\ref
{sigma2-5}) in (\ref{Eapp}) and (\ref{muapp}) we have, for $\lambda $ small
enough

\begin{equation}
E_{\varphi ^{3}-\varphi ^{5}}=E_{\varphi ^{3}}(\lambda )+\Delta E_{\varphi
^{3}-\varphi ^{5}}(\lambda ,\varepsilon ),  \label{Eng}
\end{equation}

\begin{equation}
\mu _{\varphi ^{3}-\varphi ^{5}}=\mu _{\varphi ^{3}}(\lambda )+\Delta \mu
_{\varphi ^{3}-\varphi ^{5}}(\lambda ,\varepsilon ),  \label{potG}
\end{equation}
where $E_{\varphi ^{3}}(\lambda )$ and $\mu _{\varphi ^{3}}(\lambda )$
correspond to the energy and the chemical potential, respectively, for the
cubic NLSE and are given, up to the 5th order, by~\cite{rolci}

\begin{eqnarray}
E_{\varphi ^{3}}(\lambda ) &=&1+\frac{\sqrt{2}\lambda }{4\sqrt{\pi }}-\frac{
1 }{\pi }\frac{1}{64}\lambda ^{2}+\frac{\sqrt{2}}{12\pi \sqrt{\pi }}\frac{3}{
256}\lambda ^{3}  \notag \\
&&-\frac{1}{32\pi ^{2}}\frac{1}{256}\lambda ^{4}+\frac{\sqrt{2}}{2^{13}\pi
^{2}\sqrt{\pi }}\frac{3}{2^{6}}\lambda ^{5},  \label{efi3}
\end{eqnarray}

\begin{eqnarray}
\mu _{\varphi ^{3}}(\lambda ) &=&1+\frac{\sqrt{2}\lambda }{2\sqrt{\pi }}-
\frac{1}{\pi }\frac{3}{64}\lambda ^{2}+\frac{\sqrt{2}}{\pi \sqrt{\pi }}\frac{
1}{256}\lambda ^{3}  \notag \\
&&-\frac{5}{32\pi ^{2}}\frac{1}{256}\lambda ^{4}+\frac{3\sqrt{2}}{2^{12}\pi
^{2}\sqrt{\pi }}\frac{3}{2^{6}}\lambda ^{5}.  \label{chefi3}
\end{eqnarray}
The terms $\Delta E_{\varphi ^{3}-\varphi ^{5}}(\lambda ,\varepsilon )$ and $
\Delta \mu _{\varphi ^{3}-\varphi ^{5}}(\lambda ,\varepsilon )$ tell us how
much error is made by neglecting the potential $-\varepsilon \lambda
^{2}\left\vert \psi \right\vert ^{4}$ in the 1D cigar-like shape model.
These quantities are expressed by
\begin{eqnarray}
\Delta E_{\varphi ^{3}-\varphi ^{5}}(\lambda ,\varepsilon ) &=&\frac{\sqrt{3}
}{9\pi }\left[ -\lambda +\frac{\sqrt{2}}{8\sqrt{\pi }}\lambda ^{2}\right.
\notag \\
&&\left. -\frac{1}{32\pi }\lambda ^{3}+\frac{7\sqrt{2}}{2^{11}\pi \sqrt{\pi }
}\lambda ^{4}\right] \varepsilon \lambda ,  \label{enererror}
\end{eqnarray}
and

\begin{eqnarray}
\Delta \mu _{\varphi ^{3}-\varphi ^{5}}(\lambda ,\varepsilon ) &=&\frac{
\sqrt{3}}{3\pi }\left[ -\lambda +\frac{\sqrt{2}}{6\sqrt{\pi }}\lambda
^{2}-\right.  \notag \\
&&\left. \frac{5}{96\pi }\lambda ^{3}+\frac{7\sqrt{2}}{2^{10}\pi \sqrt{\pi }}
\lambda ^{4}\right] \varepsilon \lambda .  \label{mufi2-5}
\end{eqnarray}
Figure~(\ref{2}) is devoted to the calculated chemical potential $\mu
_{\varphi ^{3}-\varphi ^{5}}$ using Eqs.~(\ref{potG}), (\ref{chefi3}), and (
\ref{mufi2-5}) as a function of $\lambda $ for $\varepsilon =0,$ 0.05, and
0.1$.$ First, by comparing Figs.~(\ref{1}) and (\ref{2}) we see the strong
qualitative difference between the two types equations here considered,
quintic and\ cubic-quintic NLSEs. Notice that this difference is remarkable
even at $\lambda \approx 0$. From the Fig.~(\ref{2}) we can assert that the
cigar-like shape approximation\ retaining term up to $\left\vert \phi
\right\vert ^{2}$ is a good approach solely for the repulsive case. The
chemical potential (and also, the energy) for $\lambda >0$ and $\omega
<<\omega _{\mathbf{r}}$ is almost independent of $\varepsilon =3\ln
(4/3)\omega /\omega _{\mathbf{r}},$ while a not negligible contribution is
reached to $\mu _{\varphi ^{3}-\varphi ^{5}}$ if the atom-atom interaction
is attractive, even for very small value of $\varepsilon .$ In the former
case the term $-\varepsilon \lambda ^{2}\left\vert \phi \right\vert ^{4}$ is
responsible for the strong obtained dispersion compared to the $\varepsilon
=0$ limit. In order to understand from the physical point of view the
behavior of the chemical potential on $\lambda ,$ we define the effective
potential
\begin{figure}[h]
\includegraphics[scale=0.38]{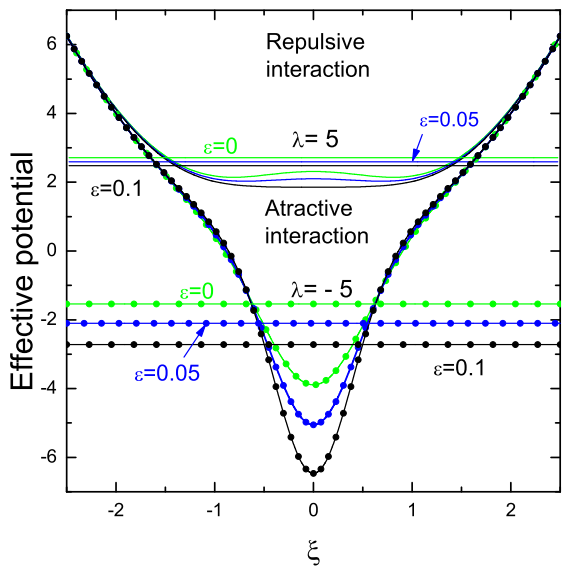}
\caption{(Color online) The potential $U_{eff}$ given by Eq.~(\protect\ref
{efectivepoten}) for $\protect\lambda =5$ (solid lines) and $\protect\lambda 
=-5$ (dot-solid lines). In the calculation we employed the values of $
\protect\varepsilon =0$ (green)$,$ 0.05(blue) and 0.1 (black). The chemical
potential $\protect\mu _{\protect\varphi ^{3}-\protect\varphi ^{5}}$ for
each $\protect\varepsilon $ is indicated by flat lines.}
\label{3}
\end{figure}
\begin{equation}
U_{eff}=\xi ^{2}+\lambda \left\vert \varphi _{\lambda }\right\vert
^{2}-\varepsilon \lambda ^{2}\left\vert \varphi _{\lambda }\right\vert ^{4},
\label{efectivepoten}
\end{equation}
where the order parameter $\varphi _{\lambda }(\xi ):=\psi _{\tau (\lambda
)}(\xi )$ has been substituted by the trial function (\ref{trialf}) with $
\sigma _{\varphi ^{3}-\varphi ^{5}}(\lambda ;\varepsilon ):=\sqrt{\tau
_{\varphi ^{3}-\varphi ^{5}}}.$ Figure~(\ref{3}) shows the potential $
U_{eff} $ for both considered cases, attractive ($\lambda =-5$) and
repulsive interatomic ($\lambda =5$) interactions. Also, in the figure is
represented the values of $\mu _{\varphi ^{3}-\varphi ^{5}}$ for $
\varepsilon =0,$ 0.05, and 0.1. It becomes clear that for a given $\lambda >0
$, the variation of the function $U_{eff}(\xi ;\lambda ,\varepsilon )$ with
respect to the parameter $\varepsilon $ is negligible$.$ Hence, the
corresponding chemical potentials $\mu _{\varphi ^{3}-\varphi ^{5}}(\lambda
,\varepsilon )\simeq \mu _{\varphi ^{3}-\varphi ^{5}}(\lambda ,0).$ Thus,
the nonlinear potential $\lambda \left\vert \psi \right\vert ^{2}$ for $
\lambda >0$ becomes a very good description to study the physical properties
of a 1D cigar-shape BEC under the condition that the transverse harmonic
oscillator frequency $\omega _{\mathbf{r}}$ is much larger than the
perpendicular frequency trap $\omega $. This result is in agreement with the
conclusions of Ref.~ \onlinecite{previsiondelamda0}. As one can see from
Fig.~(\ref{2}) the error $\Delta \mu _{\varphi ^{3}-\varphi ^{5}}(\lambda
>0,\varepsilon )$ ranges between 2-7 \% with respect to the value of $\mu
_{\varphi ^{3}}(\lambda )$. Now, if we consider the opposite case, i.e. an
attractive interaction, the behavior of $U_{eff}(\xi ;\lambda ,\varepsilon )$
presents a strong localized potential and, in correspondence, the chemical
potential will change drastically as the parameter $\varepsilon $
increases.\ This characteristic of the effective potential (\ref
{efectivepoten}) for $\lambda <0$ determines clearly that the term $\lambda
\left\vert \psi \right\vert ^{2}$ is not enough for a correct description of
1D cigar-like shape BECs. In this case, the residual three dimensionality
term $-\varepsilon \lambda ^{2}\left\vert \psi \right\vert ^{4},$ for the
effective 1D GPE under a harmonic trap$,$ strongly modify the corresponding
chemical potential. Small variation of the strength $\varepsilon \lambda ^{2}
$ leads to strong change of $\mu _{\varphi ^{3}-\varphi ^{5}}$ value and
hence the ground state energy $E_{\varphi ^{3}-\varphi ^{5}}$ as well.

This peculiar behavior is related to the orbital stability of the nonlinear
Schr\"{o}dinger equation (\ref{CQAdG}). As mentioned above, the 3D GPE
presents a set of ground states which is orbitally stable for any value of
the self-repulsive interaction, while for the attractive interparticle
interaction regime ($a_{s}<0$) the solution does not collapse if and only if
the\ condition $Na_{s}>-a_{\perp }0.627$ is fulfilled. The former criterion
can be rewritten as $-\lambda <\lambda _{3D}=2.33/\sqrt{\varepsilon }$ and
it should be compared with condition of existence and validity, for $
-\lambda <\lambda _{S}=2.857/\sqrt{\varepsilon },$ of the obtained solutions
Eqs.~(\ref{Eng})-(\ref{mufi2-5})$.$ In the inset Fig.~(\ref{2}) for the
attractive interaction, the intersection of these two sets, $\lambda
_{3D}(\varepsilon )\cap \lambda _{S}(\varepsilon )$, is represented by a
shaded region in the $\lambda -\varepsilon $ diagram.

In conclusion, a new variational approach is presented, which allows to
construct for the cubic-quintic GPE closed analytical expressions for the
order parameter, the minimal energy, and the corresponding chemical.
Focusing on the compact analytical expressions, we report the contribution
of the quintic term and the systematic error of the residual 3D contribution
to the 1D cigar-shape model for both attractive and repulsive interaction.
By the calculations the obtained chemical potential solution highlight the
strong dependence on the sign of strength $\lambda $ and the values of
quintic self-interaction parameter $\varepsilon $.

\acknowledgments C.T-G acknowledges the hospitality at the
Max-Planck-Institut f\"{u}r Physik Komplexer Systeme and thanks Alexander
von Humboldt Foundation for financial support.

\end{document}